\documentstyle[twocolumn,prc,aps,epsf,epsfig]{revtex}
 \begin{document}
 
\newcommand{\be}{\begin{eqnarray}}
\newcommand{\ee}{\end{eqnarray}}
\twocolumn[\hsize\textwidth\columnwidth\hsize\csname@twocolumnfalse\endcsname
\title{
Limits on Cosmological Variation of  Strong Interaction and Quark Masses \\ from
Big Bang Nucleosynthesis, Cosmic, Laboratory and Oklo Data
}

\author{ V.V. Flambaum$^1$ and E.V. Shuryak$^2$ }
\address{$^1$
 School of Physics, The University of New South Wales, Sydney NSW
2052,
Australia
}
\address{$^2$ 
Department of Physics and Astronomy, State University of New York, 
Stony Brook NY 11794-3800, USA
}

\date{\today}
\maketitle

\begin{abstract}
Recent data on cosmological variation of the electromagnetic
fine structure constant from  distant quasar (QSO)
absorption spectra
have inspired a more general discussion of possible variation
of other constants. We discuss variation of strong scale and
 quark masses. We   derive
the  limits on their relative change from (i)
primordial Big-Bang Nucleosynthesis (BBN); (ii)
 Oklo natural nuclear reactor, (iii) quasar absorption spectra, and
(iv)
laboratory measurements of hyperfine intervals.
\end{abstract}
\vspace{0.1in}
]
\begin{narrowtext}
\newpage

\section{Introduction}
Time variation of  major constants of physics
is an old and fascinating topic,  its discussion by many
great physicists -- Dirac as the most famous example --  
had surfaced many times in the past. Recent attention to this
issue was
caused by astronomical data which seem to suggest a variation
 of electromagnetic $\alpha$  at the $10^{-5}$ level
 for the time scale 10 bn years, see \cite{alpha}.
The issue discussed in this work is related to it, although
 indirectly. 
Instead of looking into atomic spectra and testing a stability 
of electric charge, we will discuss possible variations of 
$nuclear$ properties induced by a change in strong and
weak scales. We will not go into theoretical discussion
of why such changes may occur and how they can be
related to modification of electromagnetic $\alpha$.
Our aim is to identify the most stringent phenomenological
limitations on such a change,  at (i) a time of the order of few minutes,
 when the Big Bang Nucleosynthesis (BBN) took place, as well as
(ii) at the time of Oklo natural nuclear reactor (1.8 bn years ago),
(iii)
 when quasar
radiation has been absorbed in the most distant gas clouds
(3-10 bn years ago) and (iv) at the present time.

Mentioning relevant literature we start with the 
BBN limits on electromagnetic $\alpha$, obtained
 in  \cite{Rubinstein}. The
main  results come from variation of
late-time nuclear reactions. Because of low temperatures
and velocities involved at this stage, those reactions
 have quite significant suppression
due to Coulomb barriers, in spite of the fact that only Z=1-3 is involved.
These limits are in the following range
\be \label{em_limit}   |\delta \alpha |^{BBN}/\alpha <0.02 \ee

 In general, all
the models for  time variations of electromagnetic/weak/strong interactions
can be divided into two distinct classes, depending on whether
it originates in (i) $infrared$ or (ii) $ultraviolet$.
The former approach ascribe  variations to some hypothetical
 interaction of the corresponding gauge bosons
with some matter in Universe, 
such as  vacuum expectation values (VEVs) or  ``condensates'' 
of some scalar fields. Those typically 
 have zero momentum but can have
cosmological time dependence. We would not discuss it:
for recent example and references see \cite{VEVS}. 

We would however 
 mention few details from two recent examples of the latter approach,
 by Calmet and H.~Fritzsch \cite{Calmet:2001nu}
and Langacker, Segre and Strassler \cite{Langacker:2001td}. Their main
  assumption   is that
  Grand Unification \cite{GQW} of electromagnetic,
weak and strong forces holds {\em at any time}. Therefore, 
 a
 relation
between
all three coupling constants exists:   truly modified 2 parameters
are in this approach the {\em unification scale}\footnote{
 One might think that
if the GUT scale be used to set units,
its variation would be impossible to detect without explicit 
measurements related to gravity.
But it is not so, since the cosmological expansion itself
(which is quite important
for BBN) contains the Newton's constant (or the Plank mass) in the Hubble
 constant.  } 
$\Lambda_{GUT}$ and
{\em the value of the unified coupling} $\alpha_{GUT}$ at this scale.
Their time variation is assumed to
 propagate down the scales by the usual (unmodified) renormalization group.

If this assumption is correct, any variation of electromagnetic
$\alpha$ should be accompanied by a  variation of strong and weak
couplings as well. Specific predictions need a model, we will
mention the one discussed in \cite{Langacker:2001td}. In it
 the
 QCD scale  $\Lambda_{QCD}$
 (determined as usual by a continuation of the running coupling
constant into its -- unphysical -- Landau pole)
is modified as follows
\be \label{QCD}
{\delta \Lambda_{QCD} \over  \Lambda_{QCD}}\approx 34 {\delta \alpha
\over \alpha}
\ee

Another focus of our work is possible limits on cosmological
modifications of
{\em quark masses}.
According to Standard Model,
 they are related to electroweak symmetry breaking scale, as well as
to some Yukawa couplings $h_i$  .
In \cite{Langacker:2001td}  running of those  has been considered, with
a (model-dependent) conclusion that quark mass 
 indeed may have  a different (and stronger) change
 \be \label{mq}
{\delta m_q \over m_q}\approx 70 {\delta \alpha \over \alpha}
\ee
Large coefficients in these expressions are generic for GUT
and other approaches, in which modifications come from high scales:
they appear because weak and strong couplings run more.

If a coefficients of such a magnitude
are indeed there, at the BBN time of few minutes the QCD scale and quark masses 
 would be  modified quite a bit, if the upper
 limit (\ref{em_limit}) be  used in the r.h.s. 

The type of questions we are
trying to answer in this work are:
Do we know whether it might or might not
actually happened? Which simultaneous 
 change of strong and weak interaction
scales is or is not  observable?
What observables are the most useful ones, for that purpose?
 What are the actual limits on their
 variation which can be determined from
BBN and other cosmological and laboratory data? 

Let us repeat, that
although we use the above mentioned papers as a motivation, we do 
 not rely on any particular model. Nevertheless,
we will at the end of the paper return to these predictions
in order to see whether our limits on time variation imply
stronger or weaker effects than the electromagnetic ones.

\section{The role of heavy and strange quark masses in hadronic/nuclear observables}

Both papers just mentioned \cite{Calmet:2001nu,Langacker:2001td}
 argue for what
 we would  call the
$zeroth$ approximation to QCD modification. It assumes that
the QCD scale $\Lambda_{QCD}$ is so  dominant in all hadronic and
nuclear phenomena
 that all dimensional parameters -- hadronic masses,
magnetic moments,
energies of nuclear levels, etc -- are  to a good approximation
 simply proportional to its respective powers.

If so,
 $any$ time  variation of the overall strong interaction scale
 would $not$ change dimensionless quantities (such as
mass ratios or g-factors) which we
can observe. The absolute scale of hadronic and nuclear spectroscopy
 may change, but
even if we would be able to observe it
 from a cosmological
distances, such a modification would
easily be confused with the overall redshift.

Fortunately, this pessimistic situation\footnote{
Motivated historically by large number
of colors limit or quenched lattice QCD. Both may be  reasonable starting
approximations, which are not expected to be really accurate.} is in fact rather far from
reality. Quark masses do play significant role in hadronic/ nuclear
physics, and, if they have time modification different from that of
$\Lambda_{QCD}$, those can be detected. 

Logistically it is convenient to start with masses of heavy -- c,b,t --
quarks. They do play a role in running of the strong charge, from high 
scales down, changing beta function each time a corresponding mass
scale is passed. However, as it is well known,  those effects can be readily
absorbed in re-definition of $\Lambda_{QCD}$.

If strong coupling is the same at some high normalization point M,
between $m_b$ and $m_t$, the
 corresponding relations between $\Lambda_{QCD}$ with all
experimental quarks and $without$ heavy c,b,t quarks is as follows:
\be 
\Lambda_{without-c,b,t}=\Lambda_{QCD} ({M^2 \over m_c m_b})^{2/27}
\ee
Note that rather small powers of the masses are involved in this relation.
 In effect, we indeed
may pretend that c,b,t quarks do not exist at all, as far
as basic hadronic/nuclear physics is concerned.

The situation is completely
different with the next quark flavor we have to discuss, 
the {\em strange quark}.
It is still true, that if one fixes strong coupling $\alpha_s(k)$ at
some sufficiently high scale\footnote{It  is in fact done on the lattice, where
k is the inverse lattice spacing, typically 2-3 GeV.} and then considers
the role of non-zero $m_s$ in perturbative beta function, the effect is
negligible. (Since its scale is too low for QCD to be passed by.)

But  hadronic/nuclear masses and properties are not determined by
perturbative diagrams, leading to beta function: 
they are of course determined
by much more complicated $non-perturbative$ dynamics.
Although it is far from being completely understood,
it is clear that it does indeed depend strongly on quark masses.
In particular, the ``strange part'' of the vacuum energy density  
can be estimated, because the derivative of the vacuum energy
\be {\partial \epsilon_{vac} \over \partial m_s}=<0|\bar s s |0>\approx -1.4
\, fm^{-3} \ee is known\footnote{For definiteness, this number had come
from the QCD sum rules, which  typically correspond to operator normalization
at
$\mu=1 \, GeV$. The lattice numbers are similar, but
 normalized at inverse lattice 
spacing, typically $\mu=2 \, GeV$. Anomalous dimension of this
operator lead to small difference between two normalizations which we
ignore here.}. Thus the linear term in the strange part of the vacuum energy 
\be \label{0ss0}
\epsilon_s=m_s<\bar s s>\approx -0.2 \,\, GeV/fm^3 \ee
 is not negligible  compared to
gluonic vacuum energy 
     \cite{vacuum_energy}
\be \epsilon_g=-{(11/3)N_c-(2/3)N_f \over 128 \pi^2}<0| (g
G^a_{\mu\nu})^2|0> \\ \approx -(0.5-1)\, \, GeV/fm^3 \ee
making 20-40 percent of it. (The numerator in this expression is the
familiar coefficient of the QCD beta function, with $N_c=3,N_f=3$
being the number of colors and relevant flavors. It appears because
this expression, known as the scale anomaly to lowest order, 
has the same  origin
as the beta function itself.)

Furthermore, for the nucleon 
one finds that similar ``strange
fractions''
 of
their masses
are of the same magnitude,
 e.g.\footnote{ The reader may find discussion of 
phenomenological situation in the first paper from
\cite{sigmaterm}, while the second
contains lattice calculations of this quantity. For reference,
their conclusion the r.h.s. of (\ref{NssN}) is 1.53$\pm$ 0.07, with
the errors
being statistical 
 only.}
\be \label{NssN}
{\partial m_N \over \partial m_s}=<N|\bar s s |N>\approx 1.5 \ee
Putting into linear expansion  the strange quark mass $m_s=120-140 \,\,MeV$ one finds that
 about 1/5 of the nucleon
 mass comes from the ``strange
 term''.

 (Although in this paper we cannot go into discussion of why it is the 
case, let us make a small digression. First of all,
the reader should not be confused with the fact that only a
 very small fraction of the $energy$ of a fast moving
nucleon is due to the strange sea, as experimentally
measured partonic densities tell us. Such drastic
difference between vector-like (or chiral-even) and scalar-like (or chiral-odd)
operators is very  common feature of the non-perturbative QCD. 
Its origin is related with dominance of the instanton-induced
t Hooft interaction, see \cite{SS_98}. In short, it happens
because topological tunneling
events should necessarily involve $all$ light fermion flavors.)

 Returning to cosmology, we conclude that due to strange terms, any 
variation of quark masses would imply significant modification
of hadronic masses and other properties.
It remains a challenging task for model-builders and lattice practitioners
to establish to what extent the $O(m_s)$ part of hadronic masses is or is not
universal.

 In principle, what we would call the {\em most pessimistic scenario}
is possible,  in which $\Lambda_{QCD}$ and $m_s$
enter into {\em all} hadronic observables in one   combination
\be \Lambda_{eff}=\Lambda_{QCD}+K m_s\ee 
where K is some {\em universal} constant. If such a scenario
happens to be true, its
time modification can indeed be neutralized by a change of units,  since
{\em experiments can only measure dimensionless ratios}.

  In fact, 
when lattice practitioners
express the obtained
results
 in terms of the so called ``physical units''\footnote{Those units
are defined by
some non-perturbative observable such as (i) $\rho$- meson mass or (ii) 
the string tension, or (iii) the force between
two point-like charges at fixed distance. All of those
are measured on the lattice,
with whatever quark masses one wants to have,
and then put equal to its observable value in real world {\em by a decree}.}, the dependence 
on $m_s$ indeed tends to become weaker. 

However at the moment this is just a hint, with accuracy not better
than say 10 percent, and
there is no reason  to expect this scenario to be the case.
We mentioned such a pessimistic case
provocatively, emphasizing that 
at the moment we lack
solid theory which would explain how any particular hadronic observable
 depends on $m_s$. 
In general, {\em variation of $m_s$ alone} can 
noticeably influence strong interaction
parameters, since different quantities in general depends differently 
on it. Let us give an example.

One of the most important quantity for astronomical and laboratory
experiments is the magnetic moments of nuclei. For example, the ratio of
hyperfine splitting to molecular rotational intervals is proportional to
$\alpha^2 g$ where $g$ is defined by the magnetic  moment
\be \mu={g e \hbar \over 2 m_p c} \ee  
In zeroth  approximation as well as in the most pessimistic
scenario discussed above,
  only one dimensional parameter ($\Lambda_{QCD}$ and
$\Lambda_{QCD}+K m_s$, respectively) exist, so a 
dimensionless g factor cannot have any time variation.
But, we repeat, there is no general argument for such  approximations
to be accurate. The magnetic moment and the nucleon mass are not
directly
related to each other at the QCD level\footnote{For example, in
non-relativistic quark model the nucleon mass is approximately
3 times constituent quark mass, and quark magnetic moment is
given by ``quark magneton''. However even in this model there is also
a
binding energy and other corrections. 
 A constituent quark itself
is a complicated composite object, so one should not expect
its
magnetic moment to be exactly equal to the Dirac value related to its mass.
}, and the dimensionless derivative
\be \label{mu_mus}
\mu^N_s\equiv {\partial (1/\mu_N) \over \partial m_s}\ee
remains unknown even for the proton and neutron, to say nothing about
the composite nuclei.
If it is not the same as derivative (\ref{NssN}) for the nucleon mass, any 
time variation  of $m_s/\Lambda$  would induce
a variation of the nuclear g factors.

 The role of light quark masses is another issue, and a part of
magnetic
moments related to
the contribution of the so called ``pion cloud''  will be briefly
 discussed in section \ref{sec_lab}.

In summary,  any
dimensionless ratios should be viewed as a function of the ratio
\be g(t) = g({m_s(t)\over \Lambda_{QCD}(t)}) \ee
which for small variations is reduced to 
partial derivatives such as mentioned above.

\section{The role of light quark masses}
  Unlike strange quark mass, we have more solid theory explaining
what the effect  of 
a change of  {\em light quark masses} $m_u,m_d$   relative to that of the
strong
 scale can be.
The main focus of this work is this particular variation,
which at the end we will be able to constrain rather well.
 
Since the pion is a Goldstone boson, 
 its mass scales as a geometric mean between weak and strong
scales\footnote{
The QCD anomalous dimension of the quark mass is cancelled by the
opposite one of the quark condensate, in the GOR relation: thus
quark masses here are meant with the dependence on the normalization
point removed. The same remark should also be made about the term
with the strange quark.
(We thank T.~Dent, who reminded us that this comment is needed.)
} 
 \cite{GOR}
\be m_\pi^2 \sim (m_u+m_d) \Lambda_{QCD} \ee
Therefore the appropriate parameter 
characterizing the relative change
of the pion mass ratio to the strong scale can be defined as
\be \delta_\pi=\delta({ m_\pi \over  \Lambda_{QCD} })/({m_\pi \over \Lambda_{QCD}})= \ee
$$
{1\over 2}{\delta ({m_q\over \Lambda_{QCD}})/({m_q\over \Lambda_{QCD}})} 
$$
Because the pion mass determines the range of nuclear forces,
its modification leads directly to changes in nuclear properties.
The main question we would like to study is how such
a change in
the pion mass relative to that of other hadrons, described by a
non-zero $\delta_\pi$, 
is limited at the cosmological time when primordial nucleosynthesis
took place.

  The first limit on relative change of quark masses 
have been put in \cite{Langacker:2001td},
\be \label{npdiff}  -0.1 \, < \delta ({m_n-m_p \over T_\nu})\, <  \, 0.02 \ee
where 
  both numerical values come from the observational uncertainty of
$He^4
$ production, and $T_\nu$ is the freezeout temperature for neutrino-induced
weak processes. This effect is  however not very 
restrictive, for the following reasons:\\
 (i) The sensitivity of $He^4$
production to any variation
is itself not very impressive. (Below we will discuss 
$d$ or $Li^7$ yields which may vary by orders of magnitude and is much more sensitive.); \\
(ii) As correctly explained in \cite{Langacker:2001td}, $T_\nu$  
scales as 
\be 
T_\nu \sim v^{4/3}/M_P^{1/3}\ee
as a function of weak scale (Higgs VEV) v and Plank mass, while $m_n-m_p$
variation is expected to come primarily from $\delta(m_d-m_u)$,
with smaller electromagnetic correction. Therefore,
both numerator and denominator in (\ref{npdiff}) mostly reflect
the same physics and thus a large portion of the modifications,
if exist, would tend to cancel in this combination.

\section{Crude limits on scale variations at the BBN time}
 Before we come to specifics, let us explain our basic philosophy
in selection of the observables. To get the maximal sensitivity,
one has to focus on  phenomena which 
may vary by very large factors. Basically, there are
two sources for that, inside the dynamics which drives the BBN.
One is  {\em the Boltzmann factor}, which may at late times
reach 10 orders of magnitude or more; the other  is 
{\em Gamow factor} due to Coulomb barriers, which may reach 3-4
orders of magnitude. So naturally, production of the heaviest
primordial nuclides --
especially $Li^7$, which is sensitive to both of those -- would be most
promising
place to look.

We start however
with a preliminary discussion of
possible drastic changes when $\delta_\pi$ is of the order of several
percents, and then return to more delicate limits on scale
variation. 

The nucleosynthesis starts with the two-nucleon states. Standard
nuclear forces produce one bound state -- the deuteron, pn with
 T=0,J=1 -- and a virtual
triplet of states -- pp,pn,nn with T=1,S=L=J=0.    

If pion was lighter at the BBN time relative to its present value, 
it 
 leads to better binding. Rather dramatic effects
should have occured if the virtual
 states unbound by standard nuclear forces were
bound.

Specifically, a binding of (pp) state, with its subsequent
beta decay into the deuteron,
  could eliminate
free protons, in obvious contradiction with observations 
\cite{Dyson_Davies}.
Conditions for  a binding
of pn and nn states  have been studied recently in \cite{DF}.
These states  may add new paths to
nucleosynthesis.

 Note also, that if the pion was so much heavier
than now that the deuteron gets unbound, no primordial nucleosynthesis
could possible proceed at all.
We will not discuss those in this paper, as they are superseded by the
modification of the deuteron discussed below.

As the only exception to this rule, we briefly discuss another
important bottleneck on the way toward heavy nuclides,
the absence of the A=5 bound states. We have shown below
that with $\delta_\pi\approx<-0.052$ it can 
be bridged in a modified world.

\section{Defining two most important temperatures of the BBN}

 As we will argue throughout this work that
the most sensitive parameter in BBN remains the  binding 
energy  of the deuteron $|E_d|$  ($E_d$ is negative),
we provide a (very brief) account of BBN, with emphasis on the role
of deuterons.
 For  details of the evolution, the role
of all processes involved the reader should consult
appropriate reviews, e.g. \cite{SKM,Schramm}.

When the basic reaction producing deuterons
$ p+n\rightarrow d+\gamma $
is in equilibrium at high enough T, the density of deuterons relative
to
photons is of the order of
\be 
{n_d \over n_\gamma} \sim \eta^2 exp(|E_d|/T)
\ee
Here $\eta\approx 3*10^{-10}$ is the famous primordial baryon-to-photon ratio.
Although it remains unknown what have created it and even at
which stage of the Bing Bang it happened, we will assume it happens early 
and do not consider its variations.

So, the density of deuterium remains negligible small
 till the Boltzmann factor (needed for the photons to split d)
helps. In particular, when this factor is so  large than it can
compensate
one of the $\eta$, the deuteron fraction may be comparable to that
of p,n. This condition determine 
the first crucial temperature value, to be denoted by $T_{d}$
\be
\eta  exp(|E_d|/T_{d})\sim 1 \ee
 Note that  $T_{d}$ will be modified below together with
the value of $E_d$. At standard BBN parameters it is about 70 keV,
at which  $n_d$ reaches its maximum, see fig.1(a).

What happens after d reaches
this maximum,  
 is significant $reduction$ of the density of d
(and other species) due to several
reactions
leading  into the best bound light nuclei, $He^4$.
At such low T the equilibrium configuration would wipe out all of
other nuclides.Indeed, an 
advantage in binding of $He^4$   by more than 20 MeV the Boltzmann
factor is enormous. 

However, 
 the universe
expansion
 does not allow to reach such equilibrium.
Its rate leads to the {\em final freeze-out} of all production
reactions,
the stage when reaction rates and the Hubble expansion rates are
comparable.
The Hubble rate, according to one of the Friedman eqn.
 for flat Universe and zero cosmological constant (not important so
early anyway) is
\be  H^2\equiv (\dot R/R)^2=(8\pi/3) G_N \epsilon\ee
where $G_N,\epsilon$ are Newton's constant and mater energy density.
Ignoring numerical constants,  $H\sim T^2/M_P$, with $M_P$ being the Plank
mass,
 and ignoring
the T-dependence
of the reaction rate itself $<\sigma v>_T$, one can obtain the freeze-out 
temperature
\be \label{Tf}
T_{f}\sim 1/(\eta M_P <\sigma v>_{Tf}) \ee
Note how small parameter $\eta$ fight with large $M_P$.
For standard BBN it is
$T_{f}\approx 35 \, keV$.

\section{How variation of the deuteron binding affect the BBN }
Now we can proceed to discussion of how $modification$
of the fundamental interactions would affect the BBN yields.

Let us start with {\em the most pessimistic case}
 when {\em both} the strong scale
and quark masses are modified $identically$, so that this change
 can be eliminated from the
discussion 
by simply using  modified  units.
In such  units all reactions rates, bindings and
$T_d$ would be the same as in the standard BBN. The only
difference appears in the freeze-out time and temperature
 $T_f$. The reason is, 
as we changed the units to adjust to time varying scales,
the value of the Plank mass entering  relation
(\ref{Tf}) gets modified instead.

We can estimate crudely the effect of that modification
as follows. At the second stage of the BBN, $T_d<T<T_f$,
a decrease of d and $Li^7$ is roughly power-like 
\be f_{d,Li7}\sim T^{-a} \,\,\, a=6-7\ee
The fall is by about 2 orders of magnitude.
Using this trend we conclude that the modification of their yields
within  a factor 2 (the magnitude
of current error bars) corresponds to certain change in $T_f$, which
implies the following limits on
variation of the strong scale (relative to gravity)
\be \label{QCD_alone} 
({\delta (\Lambda_{QCD}/M_P) \over (\Lambda_{QCD}/M_P}) < 0.1 \ee

We now switch to discussion of the $relative$ change, between
quark masses and hadronic scale, leading to modification
 of the pion-induced forces  and consequently, of the
 deuteron binding energy.
We will look at modification of
$T_d$, which is so important
for BBN 
 final observable
yields. 
  
\begin{figure}[ht]
\hspace*{-5mm}
\begin{center}
\epsfig{file=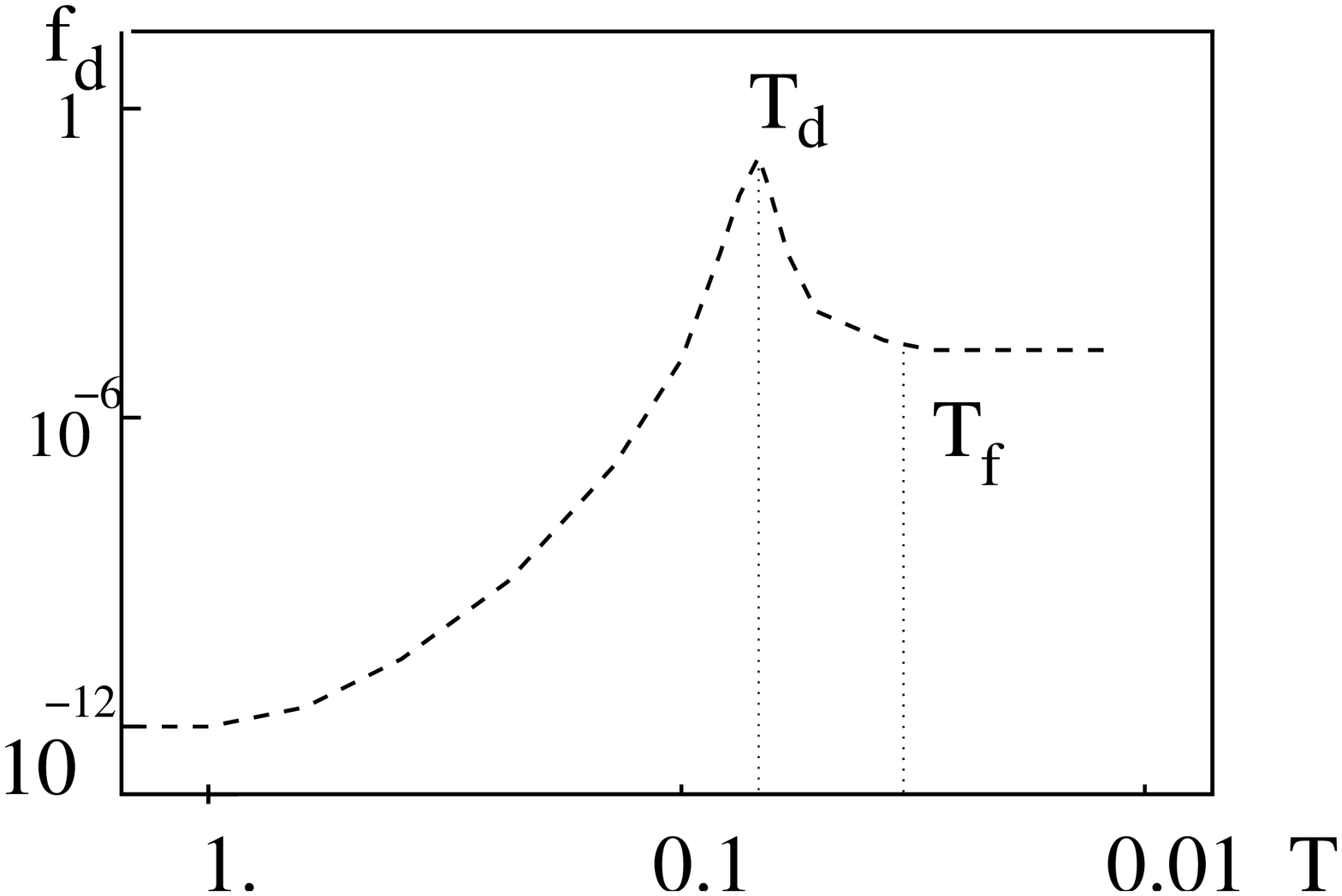, width=65mm}
\vspace*{3mm} \hspace*{-3mm}
\epsfig{file=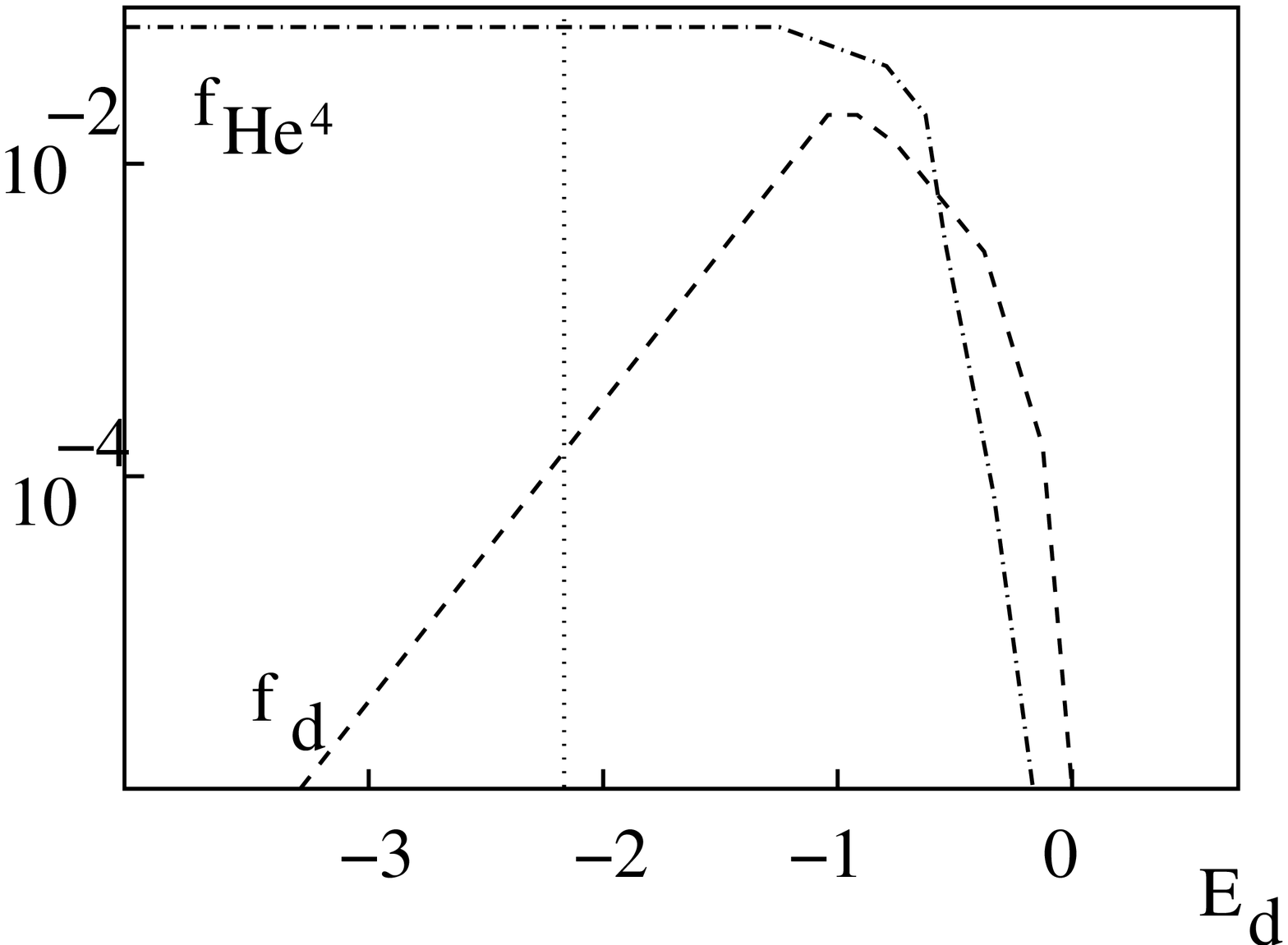, width=60mm}
\end{center}
\caption {
(a) Schematic dependence of the deuterium mass fraction $f_d$ on
temperature T (MeV). (b) Schematic dependence of the deuterium and 
$He^4$ mass fractions on the binding energy of the deuteron $E_d$ (MeV).
The vertical dotted line indicate its experimental value, -2.2 MeV.
}\label{fig:dG}
\end{figure}

A qualitative
 dependence of the 
yields of different species on the magnitude of the deuteron binding
 is schematically shown in fig.1 (b).
As $T_d$ is allowed to vary, 
the first thing to note is that there exist an {\em optimum}
 for d production corresponding to the case when
two crucial temperatures are close to each other, 
\be T_f\approx T_d\ee
It is easy to see from expressions given above, that it
 happens if the deuteron binding is around $E_d\approx -1\,  MeV$,
and its fraction reaches at this time about a percent level.

 When the value of $T_d$ reduces further, so that the
relation between the two temperatures is inverted and
 $T_f$ {\em becomes smaller  than} 
$T_d$, the stage at which d and other light nuclei 
are ``eaten'' by $He^4$  is no longer present. All heavier nuclides,
 t,$He^3$,
$He^4$ etc. reach their
 final yields from below, basically
  tracing the deuteron production.
There is no time for $He^4$ to grab most of the available neutrons
any more, and its final yield start declining (see Fig.1(b)).
Finally, when
deuterons become nearly unbound, 
 all of the primordial nucleosynthesis is wiped out altogether.

As data tell us that d fraction is at the level $10^{-4}-10^{-5}$,
the BBN happened clearly away from the maximum. Furthermore,
as $He^4$ fraction is large, we are definitely at the left
branch of the curve in Fig.1(b). This shows exponential
growth, and thus sensitivity of the results to
the variation of $E_d$ and $T_d$ can  be estimated from its
slope, which we obtained from \cite{SKM}. 
We conclude from this estimate that a change of the d yield by the
factor 2 corresponds to a constrain on relative variation of deuteron
binding
by
\be \label{lim_d}
({\delta | E_d| \over| E_d|})_d < 0.075    \ee
\section{Further limitations and the Coulomb barriers}
Now we can discuss the refinement of qualitative picture
described above, in which the absolute change of strong scale
has been eliminated by a change of units, and the only weak effect
included has been a quark mass. Now we let the electromagnetic effects
come into the game, in their simplest and
most important (exponential) form. We mean the Coulomb barriers,
described be well known Gamow factors
\be
f_{Gamow}={2\pi e^2 Z_1 Z_2 \over v} exp(- {2\pi e^2 Z_1 Z_2 \over v})
\ee
Here v is  the relative velocity, which scales as $(T/m)^{1/2}$.
As explained above, the relevant T varies in between $T_d$ and $T_f$. 
As the  former one is  modified,
 together with $E_d$, the Gamow factors
change accordingly.
 
For standard BBN we estimated that the product of Gamow factors of
reactions
leading to $Li^7$ is about $exp(14*v_0/v)$,
where $v_0$ is the unmodified value\footnote{We have also checked that
this estimate  agrees with the results
by \cite{Rubinstein} of the dependence of its yield on $\alpha$, which
come from running a compete BBN code.}.
So, assuming that experimental data on $Li^7$ are restricted within a
factor
2, we see that only a variation of 
\be \label{limLi}
 ({|\delta E_d|\over E_d})_{Li7} < 0.1 \ee 
can be 
tolerated. It  leads to a limit  on $\delta_\pi$ comparable to the one
obtained above from deuteron yield.

\section{The modified deuteron} 
In this section we try to
relate the change of deuteron binding $\delta E_d$ to the modification
of the fundamental parameter
 $\delta_\pi$ introduced above. 

In general, it is a very non-trivial dynamical
issue, which is far from being really understood.
We have discussed above to which extent the QCD vacuum energy (\ref{0ss0}) and
the nucleon mass (\ref{NssN}) depends on quark masses.   
May be one day such information
will be available from lattice QCD for nuclear forces as well, 
but right now
we do not have it and have to rely on model-dependent potentials.

Of course, one  can identify single pion exchange forces.
Especially for the deuteron channel, those lead to well known
 tensor forces producing deuteron quadrupole moment. A textbook
1-pion exchange corresponds to the following  potential
\be \label{1pi}
 V_{1\pi}={f^2 \over \tilde m_\pi^2} (\vec\tau_1\vec \tau_2)( \vec \sigma_1 \vec \partial)
(\vec \sigma_2 \vec \partial) {exp(-m_\pi r) \over r} \ee 
where $f^2=0.08$. The pion mass in denominator
has a tilde: it indicates that this mass has been put there by hand
for normalization purposes, and, unlike masses in other places,
it will $not$ have any variations.
(It has been put there in order to cancel the pion mass squared coming from 
differentiation of the exponent, which we will evaluate first.)

 When the variation of the pion mass is sufficiently small,
 the variation is
 simply given in the first order of the
perturbation theory
$
\delta E_d=<0|\delta V|0>
$
For any weakly bound state the wave function
{\em outside the potential range} can be approximated
by  simple expression
\be \label{free_WF}
\psi_0(r)=\sqrt{\kappa \over 2 \pi} exp(-\kappa r)/r
\ee 
where parameter $\kappa$ is related to binding energy by 
$E_d =  \kappa^2/m_N $.
If we use this expression till the core size, we can easily evaluate it
\be {\partial E_d \over \partial m_\pi}=-2 f^2 \frac{\kappa}{ m_\pi}
 [2 \ln{1 \over (2\kappa +m_\pi)r_0} - {m_\pi \over 2\kappa+m_\pi}]
\approx - 0.05 \ee
We therefore get from these terms
\be \label{piEd} {\delta |E_d| \over |E_d|} \approx 3 \delta_\pi  \ee
The sign tells us that reducing the pion mass we reduce binding:
this counterintuitive result comes from the $m_\pi^2$ in pre-exponent.

One more one-pion-exchange
 term we have ignored is the one in which two derivatives act on 1/r:
this leads to delta function. Naively it does not contribute since at r=0
the wave function is vanishingly small due to repulsive core. This can be
cured by the account for a finite size of the nucleon.

However the main problem with perturbative
 calculation outlined above is that the  1-pion-exchange terms
are actually several times smaller
 compared to phenomenological potentials needed
to reproduce NN scattering and the deuteron binding.
Furthermore,  
the phenomenological potentials which fit scattering
and d data actually  ascribe most of the potential 
to 2 and even 3-pion exchanges. As an example, we used 
well known work
by Hamada and Johnston \cite{HJ}, which has the central potential
in the deuteron channel in the form of subsequent pion exchanges.
We have only modified the pion mass in the exponents:

\be \label{HJ}
V(r)=V_c  Y(x)[1+a_c Y(x)+b_c Y(x)^2]\ee
\be Y(x)={e^{-x(1+\delta_\pi)}\over  x}
 \ee
where $x=r*\tilde m_\pi$ is distance in units
 of $unmodified$
pion mass, and $a_c=6$, $b_c=-1$ \cite{HJ}. The potential
also has  infinite repulsive core with
the radius 0.4 fm.

 We have ignored tensor and spin-orbit
 potentials, as having
minor
effect on the deuteron binding, adjusted the overall coefficient to
have the experimental binding energy -2.2 MeV. After that, we have
modified
the pion mass  in (\ref{HJ}), and calculated binding
energy by solving the s-wave Schroedinger eqn. For small pion modification
we need, we found good linear dependence with the following coefficient 
\be \label{modEd}
{\delta |E_d| \over |E_d|} \approx -18 \delta_\pi 
\ee
  We have presented these two estimates  as 
reasonable {\em minimal and maximal
  bounds} on this derivative. Presumably the true value is somewhere
in between 3 and -18. 

The issue has been discussed in literature. A
general discussion of how all nuclear physics 
would change if quark masses, the number of flavors
 or even the number of colors be modified, can be found in \cite{MEB}.
 Probably the latest paper addressing deuteron binding with modern
methods
is ref.
\cite{Savage_etal}. Like us, these authors
 have shown examples producing opposite sign of the derivative and conclude
that, strictly speaking, neither the magnitude nor even the sign
can be 
definitely obtained at this time. 
Further extensive lattice and  chiral perturbation theory studies are needed.

If the derivative is small  near zero, our arguments lose its weight.
However, it would be  very unnatural to get a small value of this derivative.
The attractive and repulsive parts of the potential have already
conspired
 to get
deuteron
at the binding edge. {\em One more fine tuning}, exactly at the physical
quark mass values (which does not have any special meaning
 from the QCD point of view)
is very unlikely to happen.

\section{Binding of $H\lowercase{e}^5$} 
 We now return to $He^4+n$
channel mentioned above, and evaluate which BBN limits on modification
of the pion mass
it will produce. 

In this channel there is a $p_{3/2}$ resonance
 at  energy 0.77 MeV. 
If reduction of the pion mass
can  make it bound, it will drastically enhance production of $Li$.
Without bound $He^5$ BBN has to jump over A=5, e.g. by 
a reaction $He^4+t$, which is impaired by the Coulomb barrier as well
as by very low concentration of t. 

We now study the sensitivity of $He^5$ binding 
to modifications of the nuclear potential.
Before we discuss the calculations, let us make a couple of
qualitative points.
First of all, this case is more complicated compared to deuteron,
because
we now discussed very weakly bound states, and so no simple linear
dependence on $\delta_\pi$ is expected.
Furthermore, let us point out an interesting quantum-mechanical 
 distinction between low-lying levels with zero and non-zero
angular momentum l. 
The wave function at large r, outside of the interaction range,
for zero energy level can easily be found from Schroedinger equation
\be 
\psi(r,E=0)\sim r^{-(1+l)}
\ee
In the s-wave, l=0, it is unacceptable because
such wave function is non-normalizable solution. In other words,
 keeping $exp(-\kappa r)$ in (\ref{free_WF}) is crucial, no
matter how small $\kappa$ is.
 It is no longer so for $l$=1: 
the tail of the wave function in question is now normalizable.
In short, for  $l>0$ the centrifugal barrier keeps particles 
inside the attractive potential, contrary to
 the $l=0$ states. This effect makes $l>0$ states more sensitive to the
change of the potential.

We have modeled the potential of the interaction between a neutron and
$He^4$ in the following form
\be V(r)=Q{1+w(r/c)^2 \over exp[(r-c)/\Delta]+1}\ee
where w=.445, $\Delta=.327 \, fm$. If c=1.01 fm it is a good fit to
experimental density distribution in $He^4$ \cite{Frosch}. In the potential we
simply
changed it to $c=1.01 fm + 1/m_\pi$, with variable pion mass.
The depth of the potential Q has been tuned to reproduce
the position of the above mentioned resonance, it gave
 Q=-32.6 MeV. After that we start changing the pion mass till
this level becomes bound, E=0. This happens at the magnitude
of the pion mass modification
\be \label{He}
\delta_\pi^{He5}>-0.052
\ee
If that would happen at the BBN time, the yields of Li would be
dramatically enhanced, by orders of magnitude, contrary to observations.
(This is why it is written  as the inequality.)

This value can be compared to the
natural small parameter of the virtual $He^5$ level
$E_{res}/V\sim 0.02$. The needed shift is somewhat larger because,
in spite of general argument given above, its wave function
 is spread to large r, see fig.\ref{fig_he5}.

Note that this limit  (\ref{He}) would become  an order of magnitude
weaker if we base our consideration on the direct effect of the 1-pion exchange
 eq.(\ref{1pi}) - see discussion in the previous section. For  $He^5$
there is an additional suppression  because the average value of the
potential (\ref{1pi}) over closed 1s shell is zero (the effect, however,
appears due to the exchange interaction and correlation corrections). 

As a result, we conclude that the limit from the
 $He^5$ binding cannot compete with that
from the deuteron modification in its importance in BBN. 

\begin{figure}[ht]
\hspace*{-5mm}
\begin{center}
\epsfig{file=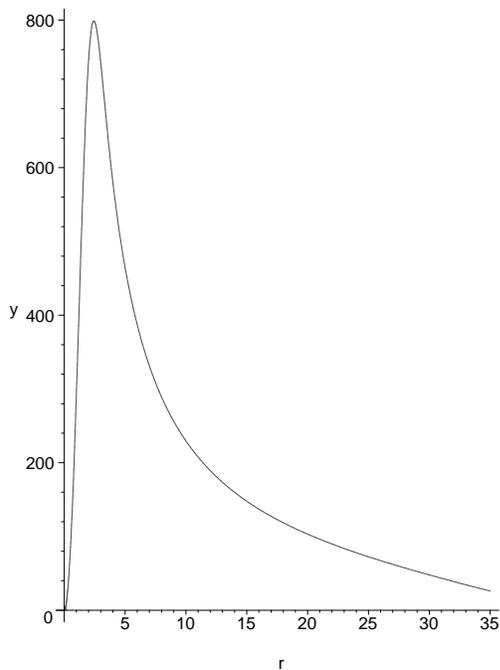, width=65mm}
\end{center}
\caption {
The wave function for zero energy neutron in $He^5$
(arbitrary normalization) versus the distance r, in
fm. 
Note that it is
normalizable,
although it does spread to rather large r. 
}\label{fig_he5}
\end{figure}

The next bottleneck, effectively blocking synthesis of heavy elements,
is the gap at A=8 nuclei, which  too readily
decay into two alpha-particles. Although we have not investigated this
case,
we do not expect it to beat the limits related to deuteron
modification as well.

\section{Limitations from Oklo data }
Finally, let us  deviate from our
discussion of BBN to a related subject,
namely similar limits following from data on natural nuclear
reactor in Oklo active about 2 bn years ago.
 The most sensitive phenomenon (used previously for limits on 
the variations of the electromagnetic $\alpha$) is disappearance of certain
isotopes (especially $Sm^{149}$) possessing a neutron resonance close to zero
 \cite{Oklo}. Today the lowest
 resonance energy is only $E_0=0.0973 \pm 0.0002 \, eV$
is larger compared to its width, so the
neutron capture  cross section $\sigma \sim 1/E_0^2$.
The data constrain the ratio of this cross section to the non-resonance
one. It
therefore implies
\footnote{Of course, under assumption that the $same$ resonance 
was the lowest one at the time of Oklo reactor.}
 that these data constrain the variation of the following
ratio $\delta( E_0/E_1) $ where $E_1\sim 1 \, MeV$ is a typical
single-particle  energy scale, which may be viewed as the energy of some 1-body
``doorway'' state.

A generic expression for the level energy in terms of fundamental parameters
of QCD can be written as follows
\be 
E_i=A_i*\Lambda_{QCD}+B_i*m_q+C_i\alpha* \Lambda_{QCD} \ee
 $$ + D_i (m_q \Lambda_{QCD} )^{1/2}+ ...
$$
where $A_i,B_i,C_i,D_i$ are some coefficients. The first term is the basic QCD
term, while others
 are corrections due to quark mass, pions and electromagnetism.
Without $B_i,D_i$ terms one can see that in the $E_0/E_1$ ratio the QCD scale drops
out, confirming our general statement above that in such kind of approximations
the variation of $\Lambda_{QCD}$ itself cannot be seen.

The sum is very small for $E_0$,
just because we deliberately picked up the lowest resonance,
 but (and this is our main point)
there is no reason to expect $each$ term to be especially small.
For example \cite{Oklo} the electromagnetic term is about 1 MeV.

Let us estimate what variation of the resonance energy would result
from a modification of the pion mass. As we did above for the deuteron,
we assume that the main effect comes from increase of the radius R of 
the nuclear potential well.
The energy of the resonances $E_i=E_{excitation}-S_n$ consists of 
excitation energy of a compound nucleus, minus the neutron
 separation energy $S_n$.
This, in turn, is a depth of the potential well $V$ minus the
neutron Fermi energy $\epsilon_F$,
$ S_n=V-\epsilon_F$. The latter scales like $1/R^2$ if the radius of the
well is changed. The kinetic part of the
excitation energy $E_{excitation}$ scales in the same way. Adding both, one 
gets shift of the resonance
\be \delta E_{i}=-(\epsilon_F+E_{excitation}){2 \delta R \over R} \ee
For resonance near zero the combination in brackets 
is approximately $V\sim 50 \,$ MeV. If
$R=5 fm+1/m_\pi$ then
 $\delta R/R=-\delta m_\pi/(R m^2_\pi)$
\be 
|{ \delta E_i \over E_i  }|= 3* 10^8 |\delta_\pi| < 0.2
\ee
The r.h.s. above comes from the observational limits claimed in \cite{Oklo}. 
The resulting limitation on pion modification
at $time\approx 1.8$ bn years ago is 
\be \label{limit_Oklo} \delta_\pi^{Oklo}<7* 10^{-10} \ee

 Note that the authors of the last work in \cite{Oklo}
found also the non-zero solution ${ \delta E_i \over E_i  } = -1 \pm 0.1$.
This solution corresponds to the same resonance moved below
thermal neutron energy. In this case 
 $\delta_\pi \approx - 4*10^{-9}$.
In principle, the total number of the solutions can be very large since
$Sm^{149}$ nucleus has millions of resonances and each of them
can provide two new solutions (thermal neutron energy on the right tale
 or left tale of the resonance). However, these extra solutions are probably
 excluded by the measurements of the neutron capture cross-sections
for other nuclei since no significant changes have been observed there
also, see \cite{Oklo}.

As in the cases of the deuteron and $He^5$ binding one may argue that
 the limits  on
 $\delta_\pi$ presented above should be weaker by an order of magnitude
since the direct contribution of  the 1-pion exchange
 eq.(\ref{1pi}) to the energy is small. To clarify this point
it may be useful to perform a numerical calculation of this
contribution to the neutron separation energy in $Sm^{150}$ and $He^5$.

\section{Limits from astrophysical and laboratory measurements}
\label{sec_lab}

Comparison of atomic H 21 cm (hyperfine) transition with molecular
rotational transitions gave the following limits on $Y\equiv\alpha^2 g_p$
\cite{Murphy1}\\ $\delta Y/Y=(-.20 \pm 0.44) 10^{-5}$ for redshift
z=0.2467
and  $\delta Y/Y=(-.16 \pm 0.54) 10^{-5}$ for 
z=0.6847. The second limit corresponds to roughly t=6 bn years ago.

As we have already discussed above, only in the
{\em most pessimistic scenario} all strong interaction phenomena 
depends on
only one parameter, e.g. $\Lambda_{eff}=\Lambda_{QCD}+K* m_s$,
its time variation cannot change dimensionless quantities like proton
magnetic
g-factor $g_p$. If so, the limits given above are just limits on
variation of $\alpha$. However, in general there is no reason to think
this to be the case, and one may wander which limits
 can be put from these data
on a cosmological variation of $m_s/\Lambda_{QCD}$.

Another issue here is a contribution proportional to light quark
masses,
$m_u,m_d$. Let us first make a qualitative point, suggesting that
their role in magnetic moments is expected to be $lager$ than in
hadronic
masses. Hadrons are surrounded by  the so called ``pion cloud'',
which have small virtual momenta p. Masses depends on it in the form
$\sqrt{p^2+m_\pi^2}$ while magnetic moments have $\vec p\times \vec r$, r being

the distance form the center. Small masses are partly compensated
by large r in the latter but not former case. 

  Model-dependent estimates support this idea. The magnitude
of the effect varies a lot between models, and as an example
we use rather conservative treatment 
by  T.Sato and S.Sawada \cite{Sato_Sawada}. It  can be seen as
a minimal estimate:
they identified the contribution
of small virtual momenta $p<\Lambda$ by using form-factors
$\sim \Lambda^2/(\Lambda^2+p^2)$, and have shown that a consistent
picture
for p,n,d and hyperon magnetic moment emerges if the cutoff is
$\Lambda\sim  m_\pi$. Furthermore, this
contribution
is shown to be basically proportional to $\Lambda^2\sim  m_\pi^2\sim m_q$.
   They found that the cloud contribution is about 1 percent
of the proton magnetic moment, but  
7.3 percents for the neutron (to be
compared to the 1 percent level for the masses). 

 We conclude that at least due to the pion cloud effect one should expect that 
the giromagnetic ratios g for nuclei have a term proportional
to $m_{light}/\Lambda_{QCD}$ contributing of the order of several
percents. 
Combining it with the contribution proportional to
 $m_s/\Lambda_{QCD}$ (which we hope does not exactly
 conspire with the effect on masses to cancel completely), we expect
overall effect at the level of 1/10 of g is of this origin.

Assuming that it has such magnitude,
that a variation of alpha and  $m_q/\Lambda_{QCD}$
do not conspire to produce observed zero, and a simplest linear dependence
\be  g_p=g_p(m_q=0)(1+q {m_q \over \Lambda_{QCD}})\ee
 we may interpret the above mentioned limits
as the following on variation of this ratio
\be \label{lim_obsrv}
|\delta({m_q \over \Lambda_{QCD}})/({m_q \over \Lambda_{QCD}})|< 10^{-4} \ee

This should be compared with the limit on $X\equiv\alpha^2 g_p m_e/m_p$ 
\cite{Cowie} $\delta X/X=(0.7\pm 1.1)*10^{-5}$ for z=1.8.
This limit was interpreted as a limit on variation of $\alpha$
or $m_e/m_p$.
It also can be viewed as a limit on variation of $m_q/\Lambda_{QCD}$.
Although few times weaker  ($\sim 2*10^{-4}$) than the limit (\ref{lim_obsrv}),
 it corresponds to higher redshift.

The limits on variation of $m_q/\Lambda_{QCD}$ can also be obtained
from  laboratory measurements of {\em ratios of hyperfine 
splittings}.
By comparison the rates of two clocks based on different atoms, say
H and $Hg^+$ \cite{clocks}, 
 we compare\footnote{Note that this comparison gives limits on
variation of $\alpha$ only due to relativistic corrections to  $Hg^+$.}
 g-factors of quite different nuclei \cite{Karschenboim}.
 In terms of the standard shell model description, this gives the
 ratio
of proton and neutron spin g-factors.  Other examples, such as using
hyperfine transition in $Cs$
as a frequency standard, would also  involve the orbital
 g-factor, with $g_l=1$. In principle, corrections to shell model include ``exchange
currents'' which also contribute to magnetic moments. All of the above
may have different dependence on $m_q/\Lambda_{QCD}$,
 and we conclude that
\be {d \over dt}\ln{ A_1 \over A_2} =K {d \over dt} \ln{m_q \over\Lambda_{QCD} }\ee
where $A_1,A_2$ are hyperfine structure constants of different atoms,
where K, a combination of derivatives (\ref{NssN},\ref{mu_mus}),
remains unknown but can be as big as 1/10 or even larger. 
Using such tentative value  and
H, Cs and $Hg^+$ measurements \cite{clocks,Cs},
 we obtain the limit on variation of $m_q/\Lambda_{QCD}$
 about $5*10^{-13}$ per year.

\section{Summary }
Combining our strongest limits on
the deuteron binding, from deuteron  (\ref{lim_d}) and $Li^7$
(\ref{limLi}),
 corresponding
to
variation of their production by a factor 2, with a relation between
modification of the deuteron binding and modification of the pion mass
 (\ref{modEd}).
Both effects suggest about the same BBN limit on the modification of the
pion mass {\em relative to strong interaction scale} $\Lambda_{QCD}$.

Using eq. (\ref{modEd}) we obtain:
\be |\delta_\pi|_{BBN} < 0.005  \ee
Eq. (\ref{piEd}) provides a more conservative limit 
\be |\delta_\pi|_{BBN} < 0.03  \ee
and we think the true limit is somewhere in between.

We have investigated other effects, such as binding of $He^5$ or
$pp,nn,np$ (S=0)
states, but found that in these cases
the needed pion modification about an order of
magnitude
larger. (It is expected, since all these states  are
 more loosely bound than the deuteron.)
If $m_s$ modification relative to strong scale are as large as our
limit on light quark modification just mentioned, it means the nucleon 
mass can be modified within $\pm 2 \, MeV$ due to strange term.   
Note also, that our limit on quark mass modification
 is  stronger than the limit
 \cite{Langacker:2001td} coming from proton-neutron 
mass difference (\ref{npdiff}). 

We also pointed out significantly weaker limits on a $simultaneous$
modification of strong scale and $m_q$ scale
at the same rate, relative to the gravity scale   
\be
{\delta (\Lambda_{QCD}/M_P) \over (\Lambda_{QCD}/M_P)} < 0.1 \ee
Limits 
on a variation of the quark mass relative to strong
scale at the $10^{-4}$ level 
 3-10 bn years ago
follows from observations of distant objects
 (\ref{lim_obsrv}),
while at the time 1.8 bn years ago the Oklo data lead to even better
limits, at the $10^{-8}-10^{-9}$ level.

  Although there is no general relation between variation of weak,
strong and electromagnetic constants, as we mention in the
Introduction it is implied by
Grand Unification 
\cite{Calmet:2001nu,Langacker:2001td}.
If one uses those (\ref{QCD},\ref{mq}), one finds that all our limits
on relative weak/strong modification are {\em much more restrictive}
 than the corresponding limits on the modification
of the electromagnetic $\alpha$. In the case of astronomical
observations,
in which variation of alpha seems to be seen, one may either
soon find the variation of g factors, or rule out  relations between
couplings based on Grand Unification idea. 

Finally, let us emphasize that our discussion
is semi-qualitative in many aspects, and
a lot of quantitative work remains to be done.
Theory-wise, the
 most straightforward thing to do is to add modifications directly
into the  
BBN  code, and get more quantitative limits.
Although there seem to be no particular problem with the standard BBN
at the moment, it is still true that
the calculated yields and observations typically differ by 1-2
standard deviations \cite{Schramm}. Therefore
it seem to be worth wile to make a global 
fit to data with unrestricted modification parameters (like our $\delta_\pi$)
and see whether zero value would or would not be the best one.

Another challenge to the theory, probably mostly lattice simulations,
is
to clarify the issue of the dependence of various hadronic parameters
on the strange quark mass $m_s$, especially how universal are the
derivatives like (\ref{NssN}) for all hadrons.

Experimental laboratory work
 and astronomical observations
of distant objects
can significantly enhance the limits available today, 
hopefully with a non-zero effect
eventually observed.

{\bf Acknowledgments} One of the authors (ES)
is supported by the US Department Of Energy, while the other (VF)
is supported by the Australian Research
Council. We are grateful to G.E.Brown, V.F. Dmitriev and V.G. Zelevinsky
for useful discussions.

\end{narrowtext}

\begin{thebibliography}{99}
\bibitem{alpha}
  J. K. Webb , V.V. Flambaum, C.W. Churchill, M.J. Drinkwater,
 and J.D. Barrow,
 Phys. Rev. Lett., 82, 884-887, 1999.
  J.K. Webb,
M.T. Murphy, V.V. Flambaum, V.A. Dzuba, J.D. Barrow,C.W. Churchill,
 J.X. Prochaska, and A.M. Wolfe,  Phys. Rev. Lett.
87, 091301 -1-4 (2001).
 M. T. Murphy, J. K. Webb, V. V. Flambaum, V. A. Dzuba, C. W. Churchill, J.
X. Prochaska, J. D. Barrow and A. M. Wolfe,
 Mon.Not. R. Astron. Soc. 327, 1208 (2001) ; astro-ph/0012419.
M.T. Murphy, J.K. Webb, V.V. Flambaum, C.W. Churchill,
and J.X. Prochaska.  Mon.Not. R. Astron. Soc. 327, 1223 (2001);
 astro-ph/0012420.
M.T. Murphy, J.K. Webb, V.V. Flambaum, C.W. Churchill,
 J.X. Prochaska, and A.M. Wolfe.
  Mon.Not. R. Astron. Soc. 327,1237 (2001); astro-ph/0012421.
\bibitem{Rubinstein}.~Bergstrom, S.~Iguri and H.~Rubinstein,
Phys.\ Rev.\ D {\bf 60}, 045005 (1999)
[arXiv:astro-ph/9902157].
\bibitem{VEVS} K.A.Olive and M.Pospelov, hep-ph/0110377 
\bibitem{Schramm} D.N.Schramm and M.S.Turner, RMP 70 (1998) 303.
\bibitem{SKM} M.Smith, L.H.Kawano and R.A.Malaney, Astr.J.Suppl. 85
(1993) 219.
\bibitem{Calmet:2001nu}
X.~Calmet and H.~Fritzsch,
arXiv:hep-ph/0112110.
\bibitem{Langacker:2001td}
P.~Langacker, G.~Segre and M.~J.~Strassler,
arXiv:hep-ph/0112233.
\bibitem{GQW} 
H.~Georgi, H.~R.~Quinn and S.~Weinberg,
Phys.\ Rev.\ Lett.\  {\bf 33}, 451 (1974).
\bibitem{vacuum_energy}
E.~V.~Shuryak,
Phys.\ Lett.\ B {\bf 79}, 135 (1978).
M.~A.~Shifman, A.~I.~Vainshtein and V.~I.~Zakharov,
Nucl.\ Phys.\ B {\bf 147}, 448 (1979).
\bibitem{SS_98}
T.~Schafer and E.~V.~Shuryak,
Rev.\ Mod.\ Phys.\  {\bf 70}, 323 (1998)
[arXiv:hep-ph/9610451].
\bibitem{sigmaterm}
J.Gasser and H.Leutwyler, Phys.Rep.87 (1982) 77.
S.~J.~Dong, J.~F.~Lagae and K.~F.~Liu,
Phys.\ Rev.\ D {\bf 54}, 5496 (1996)
[arXiv:hep-ph/9602259].
\bibitem{GOR}
M.~Gell-Mann, R.~J.~Oakes and B.~Renner,
Phys.\ Rev.\  {\bf 175}, 2195 (1968).
\bibitem{Oklo} A.I.Shlyakhter, Nature 264 (1976) 340; T.Damour and
F.J.Dyson, Nucl.Phys.B 480 (1996) 37. Y.Fujii, A.Iwamoto, T.Fukahori,
T. Ohnuki, M. Nakagawa, H. Hidaka, Y. Oura, P. Moller.
Nucl.Phys.B 573 (2000) 377.
\bibitem{HJ}
T.~Hamada and I.~D.~Johnston,
Nucl.\ Phys.\  {\bf 34}, 382 (1962).
\bibitem{Frosch}  R.F. Frosch et al. Phys. Rev. {\bf 160}, , 874 (1967).
\bibitem{Murphy1}
M.T. Murphy, J.K. Webb, V.V. Flambaum, M.J. Drinkwater, F. Combes
and T. Wiklind. Mon.Not. R. Astron. Soc. 327, 1244 (2001); astro-ph/0101519.
\bibitem{Cowie} L.L. Cowie and A. Songalia,
 Astrophys. J. {\bf 453}, 596 (1995).
\bibitem{clocks} J.D. Prestage, R.L. Tjoelker, and L. Maleki. Phys. Rev. Lett
{\bf 74}, 3511 (1995).
\bibitem{Karschenboim} S.G. Karshenboim, privite communication.
\bibitem{Cs} N.A. Demidov, E.M. Ezhov, B.A. Sakharov, B.A. Uljanov, A. Bauch,
and B. Fisher, in Proceedings of the 6th European Frequency and Time Forum.
Noordwijk, the Netherlands, 1992 (European Space Agency, Noordwijk,1992),
pp.409-414. L.A. Breakiron, in Proceedings of the 25th Annual Precise
Time Interval Applications and Planning Meeting, NASA conference
 publication No. 3267 [U.S. Naval Observatory Time Service Department (TSS1),
Washington DC, 1993], pp. 401-412. 

\bibitem{Dyson_Davies} F.J.Dyson, Scientific American 225 (1971) 1291. 
P.C.W.Davies, J.Phys.A. Gen Phys. 5 (1972) 1296
\bibitem{DF}T.~Dent and M.~Fairbairn,
``Time-varying coupling strengths, nuclear forces and unification,''
arXiv:hep-ph/0112279.
\bibitem{Savage_etal}S.R. Beane, P.F. Bedaque, M.J. Savage,
 U. van Kolck, nucl-th/010403. 
\bibitem{MEB}
H.~Muther, C.~A.~Engelbrecht and G.~E.~Brown,
Nucl.\ Phys.\ A {\bf 462}, 701 (1987).
\bibitem{Sato_Sawada} T.Sato and S.Sawada, Prog.Theor.Phys. 66 (1981)
1713.
(We thank M.Rho who brought this paper to our attention.)
\end{thebibliography}
\end{document}